\documentclass[9pt,twocolumn,twoside]{osajnl}
\journal{ao} % Choose journal (ao, aop, josaa, josab, ol, pr)

\title{Detection of polarization neutral points in observations of the combined corona and sky during the Aug.~21 2017 total solar eclipse}

\author[1*]{Frans Snik}
\author[1]{Steven P. Bos}
\author[1]{Stefanie A. Brackenhoff}
\author[1]{David S. Doelman}
\author[1]{Emiel H. Por}
\author[1,2]{Felix Bettonvil}
\author[2]{Michiel Rodenhuis}
\author[3]{Dmitry Vorobiev}
\author[4]{Laura M. Eshelman}
\author[4]{Joseph A. Shaw}

\affil[1]{Leiden Observatory, Leiden University, P.O. Box 9513, 2300 RA Leiden, The Netherlands}
\affil[2]{NOVA, P.O. Box 9513, 2300 RA Leiden, The Netherlands}
\affil[3]{Laboratory for Atmospheric and Space Physics, University of Colorado, Boulder, CO, USA}
\affil[4]{Montana State University, Bozeman, MT, USA}

\affil[*]{Corresponding author: snik@strw.leidenuniv.nl}

 \doi{\url{http://dx.doi.org/10.1364/AO.391814}}

\begin{abstract}
We report the results of polarimetric observations of the total solar eclipse of Aug.~21 2017 from Rexburg, ID (USA).
We use three synchronized DSLR cameras with polarization filters oriented at 0, 60 and 120 degrees to provide high-dynamic-range RGB polarization images of the corona and surrounding sky.
We measure tangential coronal polarization and vertical sky polarization, both as expected.
These observations provide detailed detections of polarization neutral points above and below the eclipsed Sun where the coronal polarization is cancelled by the sky polarization.
We name these special polarization neutral points after Minnaert and Van de Hulst.
\end{abstract}

\setboolean{displaycopyright}{true}

\begin{document}

\maketitle

\section{Introduction}
Solar eclipses and atmospheric optics effects rank highly among the favorite subjects of both professional and amateur observers of light and color in nature.
In both cases, polarization features are very prominent, and several commonalities are readily observed.
Due to Thomson scattering, the solar corona exhibits an essentially exclusively tangential (i.e.~along to a circle centered on the Sun) linear polarization pattern \cite{Vorobiev2020}.
Imaging polarimetry is widely used for studying the solar corona, either though natural or artificial eclipses \cite{Frazin2012}.
In a very similar way, Rayleigh scattering produces tangential polarization of the cloud-free sky.
Also, for both the coronal and the sky polarization, dust/aerosols can decrease the fractional linear polarization, and induce spectral slopes \cite{VandeHulst1950,vanHarten2014}.
The cloud-free sky, however, is not strictly tangentially polarized: due to multiple scattering and surface albedo effects, so-called polarization neutral points are observed above and below the Sun and the anti-solar point \cite{meteooptik,Horvath:02, Berry_2004}.
These neutral points appear where the weak tangential polarization due to small-angle single scattering is cancelled by the predominantly vertical polarization of multiply scattered light.
At even smaller angles, the polarization is vertical.
The polarization neutral points above and below the Sun are named after Babinet and Brewster, respectively, and the one above the anti-solar point is named after Arago.
The fourth neutral point, located below the anti-solar point, was predicted in \cite{colorlight}, and has since been observationally confirmed by \cite{Horvath:02}.

During a total solar eclipse, the illumination and polarization of the sky is exclusively determined by multiple scattered light that originates from beyond the edge of the Moon's shadow (single-scattered light from the faint corona is negligible) \cite{Shaw:78, Konnen:87}.
This implies that the sky polarization pattern during totality is completely different from regular sky polarization.
Along the path of totality a polarization neutral point is observed near zenith \cite{Shaw:75, Pomozi2001, Horvath:03, Sipocz:08, Shaw2019, Eshelman:2020}. 
{When the Sun/Moon themselves are not near zenith, this neutral points acts as a singularity of index $+\frac{1}{2}$ \cite{Berry_2004} for the surrounding sky polarization, i.e.~the polarization direction rotates by 180$^\circ$ upon traversing around the singularity.}
The polarization near the Sun is typically vertical, whereas it is predominantly parallel to horizon behind the observer \cite{Horvath:03, Sipocz:08,Shaw2019,Eshelman:2020}.
The detailed features of the sky polarization during an eclipse depend on the shadow geometry, on atmospheric features like clouds of water or aerosol, and on albedo properties in all directions around the Moon's shadow.

Here, we report detailed observations of the combination of coronal polarization and the polarization of the surrounding sky.
Such neutral points were predicted by \cite{Molodensky1996}, and (hints of them) also observed for earlier solar eclipses \cite{Koutchmy1977, Horvathbook, Molodenskii2009, Merzlyakov2019}.
Much in the same fashion as the occurrence of polarization neutral points in the cloud-free sky due to the cancellation of the tangential polarization due to single Rayleigh scattering by the vertical  multiple-scattering component, we observe polarization neutral points where the tangential coronal polarization due to Thomson scattering is cancelled by the multiply scattered sky polarization.
In tribute to Marcel Minnaert (1893--1970; founding father of the ``light and color in nature'' community) and Henk van de Hulst (1918--2000), who have both made seminal contributions to observations and understanding of the solar corona and its polarization \cite{Minnaert1930, VandeHulst1950}, and to describing atmospheric scattering phenomena \cite{Minnaert1968, Minnaert1974, VandeHulst1948, VandeHulst1971, VandeHulst1980}, we take the liberty to name these special polarization neutral points above/below the Sun during a solar eclipse to Minnaert and Van de Hulst, respectively.

\section{Experimental setup} 

\subsection{Instrument}

\begin{figure}
    \centering
    \includegraphics[width=\linewidth]{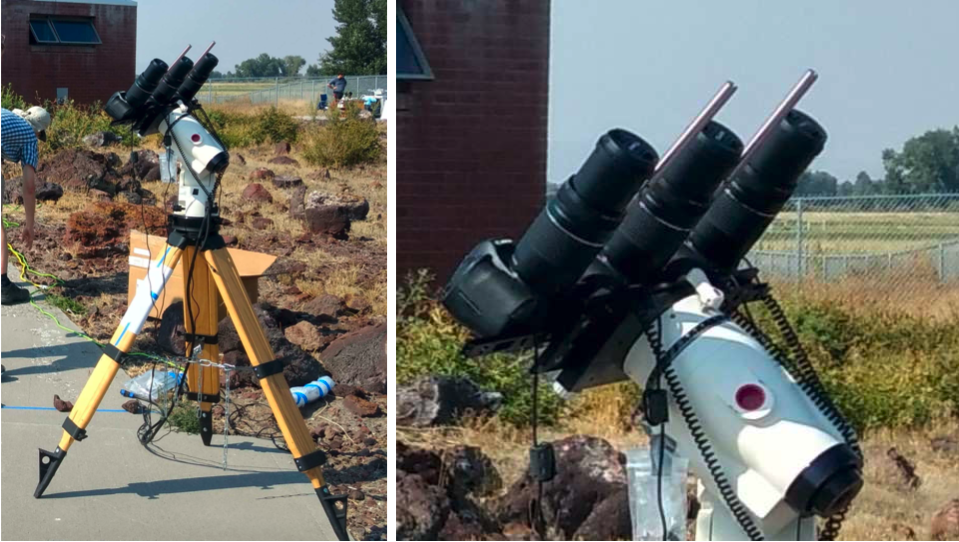}
    \caption{Photo of the LEIPo instrument with protective lens caps instead of the polarization filters.}
    \label{fig:leipo}
\end{figure}

Especially for these eclipse observations we have constructed the LEIPo instrument (Leiden Eclipse Imaging Polarimeter), consisting of three Canon EOS 6D color DSLR cameras, see Fig.~\ref{fig:leipo}.
These cameras carry 20.2 (effective) megapixels, each consisting of a 2$\times$2 RGBG Bayer pattern.
We equipped the cameras with Canon EF 75--300mm F4.0--5.6 lenses to cover a middle ground between the fisheye lenses of the co-located experiments described in \cite{Shaw2019, Eshelman:2020} and telescopic observations.
We have locked the lenses at $\sim$200 mm (at F/5.7) to obtain a field-of-view of $\sim$10$\times$7$^\circ$.
The three cameras were fixed on a common mount, that, in turn, was mounted on an amateur telescope mount.

In front of each of the three lenses we mounted a Tiffen digital HT titanium 58 mm circular polarizer, which is by design oriented such that it filters linear polarization, and converts that to circular polarization in order to not confound cameras that use an internal polarizing beam-splitter to feed a light meter.
The three effectively linear polarizers were manually oriented to roughly 0, 60 and 120$^\circ$ to provide optimal modulation for linear polarization with the minimum number of measurements \cite{delToroIniesta:00}.
With a large polarizer sheet (Edmund Optics Visible Linear Polarizing Laminated Film \# 45-204) attached to the common mount in front of all three cameras at specific predefined orientations modulo 30$^\circ$, we co-aligned all three polarizers to a common coordinate system by sequentially finding the crossed orientation for each lens.
The photographic polarizers were consecutively locked down using nail polish.
Because of the limited accuracy of this procedure, we self-calibrated the polarizer orientations during the data reduction (Sect.~\ref{sec:poldata}).

The day before the eclipse we have oriented the telescope mount with the aligned set-up towards the Sun at the time of the eclipse the next day. 
That night we attempted to manually focus all lenses at infinity by observing stars.
The next morning we checked the focus at daytime temperatures by temporarily aiming at landscape features at the horizon.
Nevertheless, both the exact focal lengths and the focus positions between the three lenses were still somewhat different, we also took this into account during data reduction (Sect.~\ref{sec:poldata}).
The system was then pointed towards the location of the Sun during the imminent eclipse at $\sim$48$^\circ$ elevation, and then remained fixed.

Before totality, we covered the instrument with a heat blanket with the reflecting side outward to prevent (over)heating. 
The cover was removed one minute before totality, and the camera exposures started with bracketed exposures (i.e.~multiple exposures with all other setting the same) 1/1000s, 1/250s, 1/60s, 1/15s, 1/4s, 1s, and 4s, at 200 ISO.
This resulted in 17 exposure sets with a cadence of $\sim$7.5 s during totality (which lasted for 2 m 17 s).
The start of each bracketed exposure of all three cameras was synchronised using a THEBEN programmable GPS controlled timer switch connected to the three camera remote-control terminals. 
All individual exposures were time-stamped using the camera's GPS system.

Back in the lab in Leiden, the RGB profiles of all three cameras were measured, using procedures outlined in \cite{Burggraaff:19}, and they were found to be sufficiently similar in behavior.
Furthermore, the dark+bias reference frames were collected for the relevant exposure times at comparable temperatures.
Finally, extinction ratios of the polarizers were measured by observing an unpolarized source (clouds) with an high quality wiregrid polarizer (WP25M-VIS from Thorlabs) in front of the camera. 
The wiregrid was rotated in steps of $5^{\circ}$ between measurements while the LEIPo polarizer remained fixed. 
The minimum and maximum intensity were subsequently determined by fitting a paraboloid to peak and valley of resulting intensity sinusoid. 
The extinction ratio follows from the ratio of the minimum and maximum and was determined to be $0.2\%$ for R, $0.03\%$ for G, and $0.7\%$ for B. 

\subsection{Location and conditions} 
Our observations were recorded at the BYU-Idaho Physics Department astronomical observatory near Rexburg, ID (USA) located at 43$^\circ$49'45.5'' N, 111$^\circ$53'06.7'' W.
Totality occured from 18:33:14 UTC to 18:35:31 UTC (11:33:14--11:35:31 local time).
This experiment was co-located with the all-sky observations described in \cite{Shaw2019, Eshelman:2020}.
During totality, the sky was completely free of clouds, with temperatures around 17.1--17.6 $^\circ$C.
Even though wildfires were rampant throughout the area in the preceding days, no clouds of smoke were reported on the day of the eclipse. Additional details about the conditions are available in \cite{Eshelman:2020}.

\begin{figure*}
    \centering
    \includegraphics[width=0.78\paperwidth]{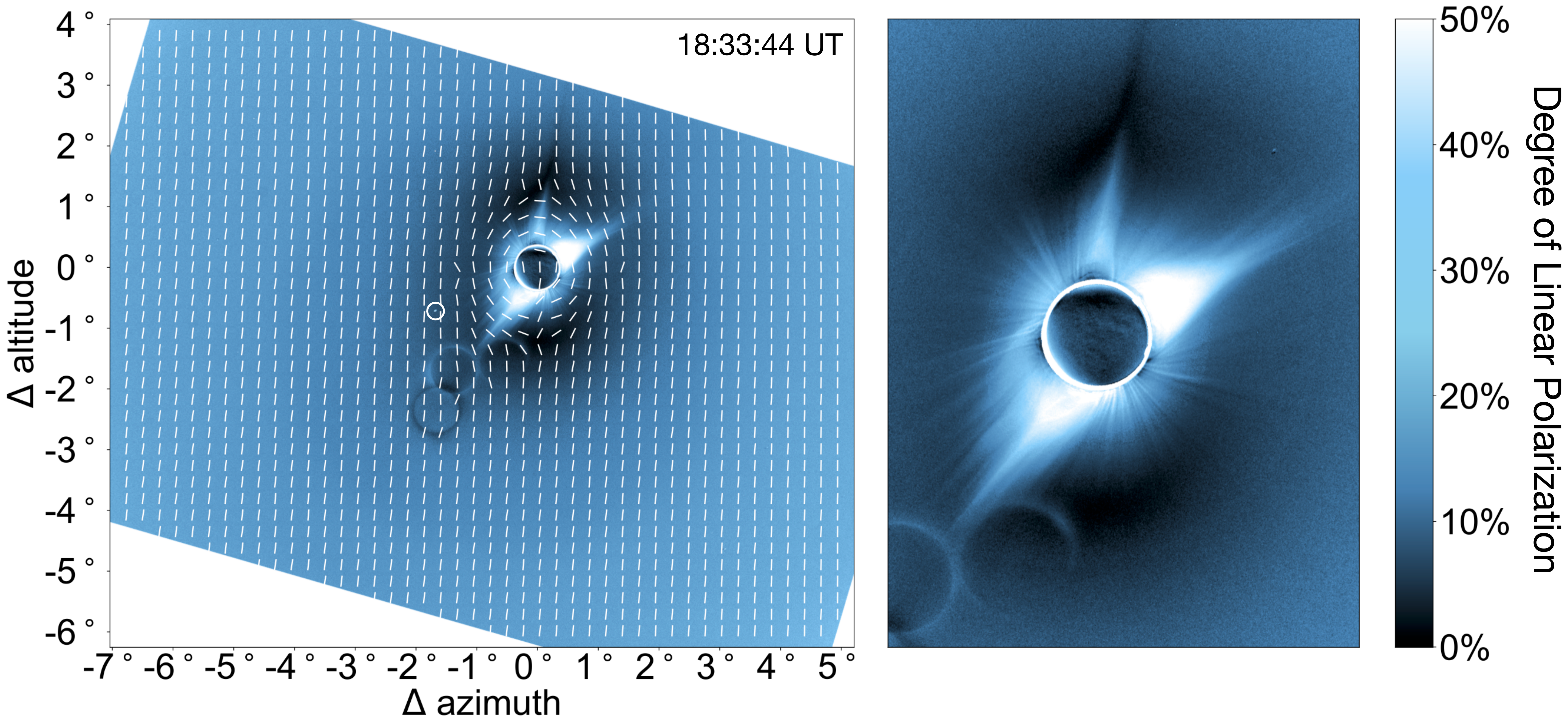}
    \caption{Degree and angle of linear polarization for a single high-dynamic-range combination of frames during totality, in the G band. The polarization vectors are plotted in the left panel in a sparse fashion to not overcrowd the imagery in degree of linear polarization. The circle indicates the position of the bright star Regulus, that exhibits spurious linear polarization with spatial structure that averages out in fractional polarization, but not in polarization degree which is necessarily $\ge$0. The zoomed-in polarization data in the right panel shows the detailed coronal structure.}
    \label{fig:coronaskypol}
\end{figure*}

\section{Data analysis}\label{sec:poldata}
The LEIPo instrument is a (fairly ad-hoc) polarimeter of the ``co-boresighted'' type \cite{Tyo:06}, that consists of complete separate optical systems with polization filters which data is to be combined.
This polarimetric implementation is known to suffer from a number of systematic effects that can induce spurious polarization signals \cite{SnikKellerreview}, and that therefore need to be addressed during data-reduction:
\begin{itemize}[noitemsep,topsep=0pt]
    \item Differential alignment (pointing, rotation) between the camera systems;
    \item Transmission / quantum efficiency differences between the camera systems;
    \item Systematic differences of exact exposure times;
    \item Synchronization errors or latencies;
    \item Different optical performance (focus, ghosting, scratches and dust specs) between the lenses;
    \item Different detector properties (e.g.~dark+bias, RGB responses, bad pixels);
    \item Different extinction ratios of the separate polarizers;
    \item Random errors to the orientations of the separate polarizers.
\end{itemize}
Because the polarizers are the first optical elements, and they are hit at (near-)normal incidence, instrumental polarization should be essentially absent.

After dark+bias subtraction and applying thresholds for low signal and overexposure, we combine each bracketed exposure set from each camera to obtain high-dynamic-range intensity images.
We assess the focus of all three cameras based on the bright star Regulus in our field-of-view, and, where necessary, degrade the image quality for the two best-focused camera systems to match that of the worst one, by a convolution with an appropriate normalized two-dimensional Gaussian.
Next, we identify co-temporal frames from the three cameras, and co-align the three data-sets during totality on a single coordinate system by sub-pixel scaling, shifting, and rotating the frames for the other two cameras, based on the centroid of the dark lunar disk (as found from a Hough transform on the inner edge of the bright corona), and the centroid position of Regulus.
To obtain polarization measurements, we treat each set of exposures from all three camera as input for a linear framework of polarimetric demodulation \cite{delToroIniesta:00}.
For an initial result, we assume that the three polarizers are perfect, and oriented exactly at 0, 60 and 120$^\circ$, and that the combination of transmission and exposure time is identical for the three cameras.
The 3$\times$3 demodulation matrix to convert three intensity measurements into the first three Stokes parameters (intensity, linear polarization at 0-90$^\circ$, and linear polarization at $\pm$45$^\circ$ \cite{Tyo:06, SnikKellerreview}) is then readily obtained as the inverse from the modulation matrix that describes the linear polarization filtering of the three polarizers \cite{delToroIniesta:00}.
To correct for transmission/exposure differences, we consider the obtained fractional linear polarization for a sufficiently large photometric aperture centered on the images of Regulus, which for our purposes can be assumed unpolarized ($\sim$4$\cdot$10$^{-5}$ \cite{Cotton+2017}), and is much brighter than the sky.
We subtract values of sky flux and polarization from a nearby aperture of equal size.
We first add two variable parameters to our modulation/demodulation model that describe the relative transmissions of two cameras, and optimize their values to force the obtained polarization of Regulus to zero.
We thus determine first estimates of the relative transmission ratio offsets in R, G and B bands.
Consequently, we add three variables for the orientation offsets of the polarizers, and obtain appropriate values for those for forcing the coronal polarization to be strictly tangential.
Finally, we apply a joint optimization of all five parameters described above, with different relative transmissions for the three color filters.
We find relative transmission ratios of up to 9\%, and angular offsets of 1--10 degrees for the three polarizers, which reflects our fairly haphazard manual alignments.
In the corona, we find root-mean-square offsets in polarization angle with respect to pure tangential of up to 2.5$^\circ$, which we attribute exclusively to residual systematics, and not to the coronal structure itself \cite{Vorobiev2020}.
Furthermore, we apply the corrections for the finite extinction ratios of the polarizers at the different color bands.
We estimate an absolute accuracy for fractional polarization / polarization degree of a few percent, and an accuracy for polarization angle of a few degrees, and we note that the polarimetric performance can vary between the R, G and B filters.

With the calibration images of fractional linear polarization, we perform a polarimetric high-dynamic-range combination by adopting the median value of fractional polarization at each pixel with intensity thresholds applied to the bracketed exposure source frames.
Finally, these images of fractional linear polarization are converted to degree and angle of linear polarization.

\section{Results}
Fig.~\ref{fig:coronaskypol} shows a frame of degree and angle of linear polarization in alt-az coordinates (i.e.~the principal plane is vertical and goes through the centroid of the Sun).
Even though we do not plot the polarization vectors in the densest possible way, it is clear that the corona is tangentially polarized.
The zoomed panel on the right shows the polarized coronal structure.
The Minnaert and Van de Hulst polarization neutral points are clearly visible as black regions of low polarization degree above and below the Sun, although they obtain an extended appearance due to the structure of the corona.

Three ghost images of the inner corona are clearly visible below the Sun, and show up as false polarization signals as the three separate camera systems were not exactly co-aligned, and the intensity ghosts propagate in different locations through the polarimetric demodulation procedure.
Also the bright ring that appears at the edge of the lunar disk is false, and is cause by imperfect co-alignment.
From the lunar limb outward the coronal polarization first increases and then decreases, cf.~the model first put forward by Van de Hulst \cite{VandeHulst1950, Vorobiev2020}.
The maximum observed polarization degree in the corona is 56\% for the G color band, which matches the values presented in \cite{Vorobiev2020}.

The sky polarization surrounding the corona, and also in front of the lunar disk is vertical to within 1.1$^\circ$around mid-totality, and uniform (RMS=3.1$^\circ$).
During totality, this sky polarization rotates by 6.9$\pm$4.4$^\circ$.
In Tab.~\ref{tab:DoLPsun} we list the measured degree of linear polarization at the position of the Sun (DoLP($\odot$)), and measured at $\Delta$(az,alt) [-4$^\circ$,3$^\circ$] next to the Sun as defined in Fig.~\ref{fig:coronaskypol}, and compare with the co-located all-sky observations presented in \cite{Shaw2019, Eshelman:2020}, and the telescopic observations of \cite{Vorobiev2020} from Madras, OR.
As is readily visible in Fig.~\ref{fig:coronaskypol}, the measured sky polarization in front of the Moon is generally much weaker than the surrounding sky polarization (although the polarization angle is consistent).
In Tab.~\ref{tab:DoLPsun} we also compare the sky polarization sampled at $\sim$5$^\circ$ next to the Sun with the all-sky observations presented by \cite{Shaw2019, Eshelman:2020}.
From the telescopic observations in \cite{Vorobiev2020} it is impossible to derive the sky polarization far away enough from the Sun such that the corona does not dominate the signal.
The intrinsic measurement uncertainties are large, as they are dominated by photon and read noise.
Furthermore, the uncertainties in Tab.~\ref{tab:DoLPsun} do not contain systematics of the two polarimetric systems that both consist of three co-boresighted DSLR cameras.
Considering absolute polarimetric accuracies of a few percent in combination with the measurement noise and the fact that two samples are not taken at the exact same sky position and time during the eclipse, the two separate observations at Rexburg are mutually consistent.
Because of the relatively large measurement uncertainties, we do not draw any conclusions about polarization color effects.

\begin{table}
\begin{tabular}{ |l|l|l|l|l| }
 \hline
 & \cite{Shaw2019, Eshelman:2020} & this work & this work &  \cite{Vorobiev2020} \\ 
 & Rexburg, ID & Rexburg & Rexburg & Madras, OR \\
& DoLP($5^\circ$) & DoLP($5^\circ$) & DoLP($\odot$) & DoLP($\odot$) \\
  \hline
 B & 0.33$\pm$0.07 & 0.26$\pm$0.01 & 0.11$\pm$0.01 & 0.16$\pm$0.01 \\ 
 G & 0.29$\pm$0.04 & 0.25$\pm$0.01 & 0.09$\pm$0.01 &  \\
 R & 0.32$\pm$0.04 & 0.24$\pm$0.02 & 0.04$\pm$0.02 & 0.08$\pm$0.01 \\ 
 \hline
\end{tabular}
     \caption{Comparison of degree of linear polarization (DoLP) of the sky at and near the  position of the Sun+Moon for different observations and sites. The uncertainties quoted are the 1$\sigma$ for a cluster of pixels, and do not include any overall systematics.}
    \label{tab:DoLPsun}
\end{table}

\begin{figure*}
    \centering
    \includegraphics[width=0.8\paperwidth]{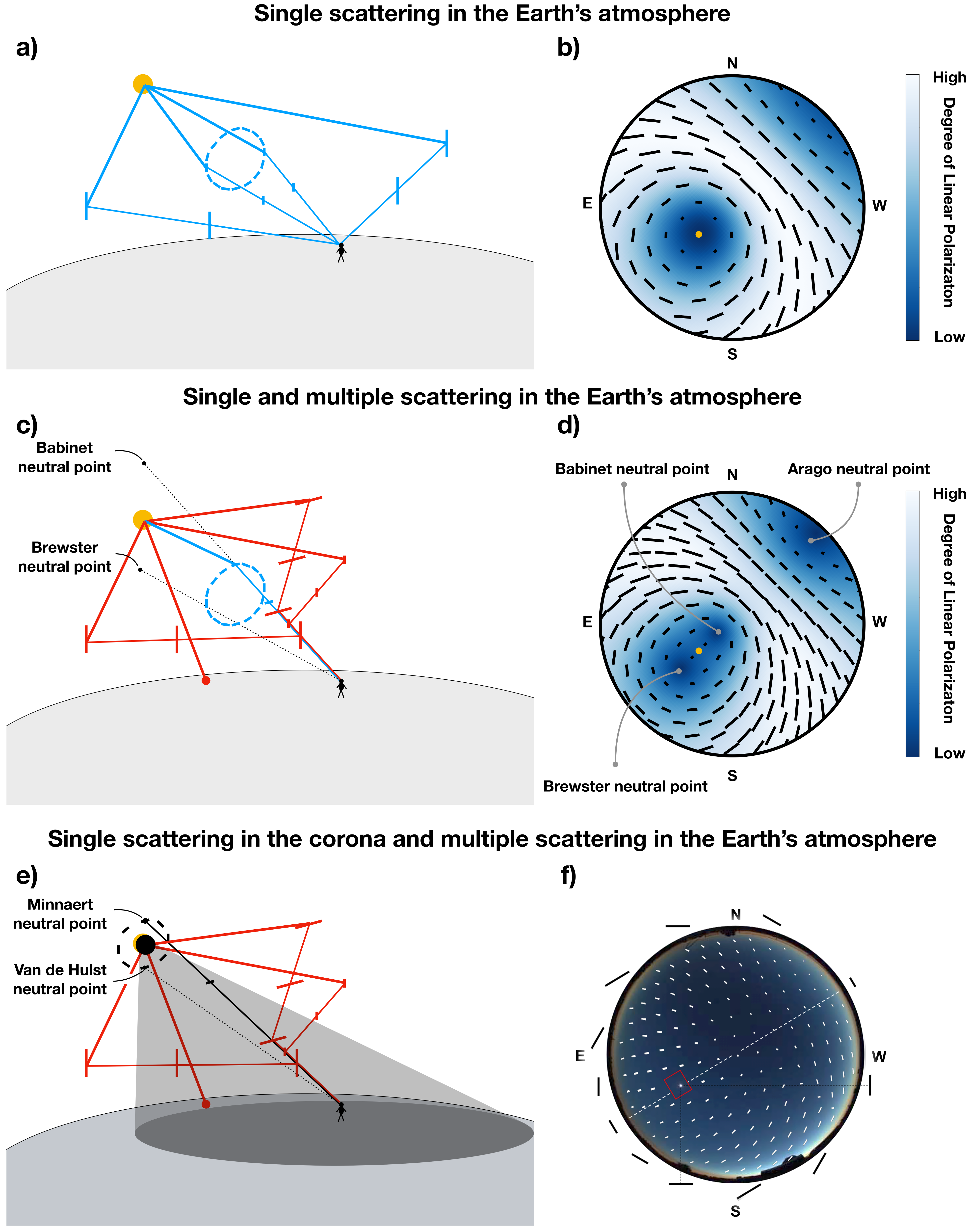}
    \caption{Qualitative explanations of sky polarization patterns of cloud-free sky outside of eclipses and during a total solar eclipse. Panel a shows single scattering (in blue lines) giving rise to the sky polarization pattern in panel b (cf. \cite{Berry_2004}). The additional multiple scattering (red lines) in panel c causes a vertical sky polarization around the Sun and the anti-solar point, giving rise to the polarization neutral points in panel d. In the case of a total solar eclipse (panel e), only the multiple-scattered component  is left, and creates a sky polarization as in panel f (data adapted from \cite{Shaw2019, Eshelman:2020}). In combination with the tangential coronal polarization (black vectors), the Minnaert and Van de Hulst neutral points emerge. The red rectangle in panel f) denotes the field of view of Fig.~\ref{fig:coronaskypol}. }
    \label{fig:cartoons}
\end{figure*}

\section{Discussion \& Conclusion}
Although a full quantitative analysis by means of a radiative transfer simulation is beyond the scope of this paper, we provide a qualitative explanation for the observed behavior of the sky polarization in combination with the coronal polarization.
The cartoons in Fig.~\ref{fig:cartoons}a and c show the single and multiple scattering behavior, respectively, for a cloud-free sky.
For single scattering (indicated with blue lines), a pure tangential polarization is observed (see Fig.~\ref{fig:cartoons}b).
Because of multiple scattering (red lines in Fig.~\ref{fig:cartoons}c) and symmetry-breaking by absorption at the Earth's surface because of albedo effects, a predominantly vertically polarized component is added in the direction of the Sun and the anti-solar point.
This gives rise to the well-known polarization neutral points at the locations where this vertical polarization cancels the weak tangential polarization above and below the Sun and the anti-solar point (see Fig.~\ref{fig:cartoons}d).
During a total solar eclipse, the single-scattering component is removed by the Moon's shadow, and the observed sky only consists of multiple-scattered light originating from outside the shadow, see Fig.~\ref{fig:cartoons}e.
This all-sky polarization pattern has been observed during at least three different solar eclipses \cite{Horvath:03, Sipocz:08, Shaw2019, Eshelman:2020}, all at locations with the Sun+Moon at intermediate altitude angles in the sky.
Also for the 2017 solar eclipse observed from Rexburg, the sky polarization was indeed predominantly vertical at the location of the Sun, and horizontal on the other side of zenith along the principal plane \cite{Shaw2019, Eshelman:2020}, see Fig.~\ref{fig:cartoons}f.
In Fig.~\ref{fig:cartoons}f we sketch a qualitative explanation for this sky polarization pattern, following \cite{Pomozi2001}. 
Beyond the horizon (and the lunar shadow) the single-scattered polarization angle and degree are indicated, and for every point in the sky it is readily deduced how the multiple-scattered sky polarization is dominated by the single-scattered polarization at the nearest horizon.
Note, furthermore, that the observed pattern \cite{Shaw2019,Eshelman:2020} close to the horizon furthest from the Sun becomes vertical again, similar to the multiple-scattered component for regular cloud-free sky that causes the Arago point around that location.
The vertical sky polarization around the Sun, together with the strong and tangential coronal polarization, makes neutral points inevitable above and below the Sun during an eclipse.
The physical origins of the four regular polarization neutral points (Babinet, Brewster, Arago and the fourth one \cite{Horvath:02}) and the two special Minnaert and Van de Hulst neutral points that were introduced in this paper, and that can only be observed during a total solar eclipse, are therefore very similar:
They are both caused by the multiple-scattered component of sky polarization cancelling out a single (Rayleigh/Thomson) scattered tangential polarization pattern.
The main difference in the polarization pattern near the Sun is that for a regular cloud-free sky, the polarization is vertical close to the Sun and tangential further away, while for the eclipse case it is the other way around.

We attribute the weak measured sky polarization in front of the Moon to (atmospheric/instrumental) scattered light from the surrounding tangentially polarized corona that adds signal that is only weakly polarized on average.
In fact, this effect also creates artificial polarization neutral points within the lunar limb given sufficient spatial resolution \cite{Vorobiev2020}.
This also means that it is difficult to compare the measurements of sky polarization at the position of the Moon between the observations presented here and those in \cite{Vorobiev2020}, as the spatial resolution and also the scattered-light properties of the instruments are different.
The fact that the values for DoLP($\odot$) that we find are smaller than those in \cite{Vorobiev2020} and inconsistent with those in Tab.~\ref{tab:DoLPsun}, could be explained by the fact that we have less spatial resolution, and therefore more polarization averaging.
Also the atmospheric and surface conditions at Madras were quite different from Rexburg, with some nearby smoke clouds that could induce more aerosol scattering, and more desert-like albedo effects, although both effects would intuitively decrease the sky polarization degree.

We have demonstrated here that a relatively low-cost and low-tech instrument can still deliver high-quality polarimetric observations.
These observations provide detailed imagery of the combined coronal and sky polarization, and their interaction to form the Minnaert and Van de Hulst neutral points.
The exact location of these neutral points obviously depend on the coronal structure, atmospheric scattering properties (including aerosol concentrations and structures), and the geometry of the Moon's shadow through the atmosphere and on surface albedo features.
Nevertheless, these neutral points are very likely a universal feature, as the polarized flux of the sky will always become brighter than the coronal polarized flux at some angular distance from the Sun.
In most cases, the sky polarization surrounding the Sun will be roughly vertical, such that the neutral points will appear above and below the Sun.
However, for a total solar eclipse observed near zenith, the sky polarization is likely to become tangential for symmetry reasons, and exhibit a neutral point of index +1 close to the location of the Sun.
In that case the combined polarization pattern will be mostly tangential, with possibilities of complicated local neutral point topologies.

Although the Minnaert and Van de Hulst polarization neutral points are related to aerosol scattering, we do not foresee any practical applications of the observations described here, as these neutral points originate from completely disjunct sources of polarization.
Yet, we are convinced that they are a unique example of the synthesis of two popular manifestations of light and color in nature: polarized scattering in the solar corona and polarized scattering in the Earth's atmosphere observed at the right location at the right time during a total solar eclipse.

\section*{Acknowledgements}
We thank the two anonymous referees for valuable comments. 
We thank Herv\'e Lamy for lending us two Canon cameras.

\section*{Funding}
European Research Council (ERC) StG (678194; FALCONER).

% Bibliography
\bibliography{Snik+2020-eclipse_corona+sky_polarization}

% Full bibliography added automatically for Optics Letters submissions; the following line will simply be ignored if submitting to other journals.
% Note that this extra page will not count against page length
%\bibliographyfullrefs{sample}

%Manual citation list
%\begin{thebibliography}{1}
%\bibitem{Zhang:14}
%Y.~Zhang, S.~Qiao, L.~Sun, Q.~W. Shi, W.~Huang, %L.~Li, and Z.~Yang,
 % \enquote{Photoinduced active terahertz metamaterials with nanostructured
  %vanadium dioxide film deposited by sol-gel method,} Opt. Express \textbf{22},
  %11070--11078 (2014).
%\end{thebibliography}

% Please include bios and photos of all authors for aop articles
\ifthenelse{\equal{\journalref}{aop}}{%
\section*{Author Biographies}
\begingroup
\setlength\intextsep{0pt}
\begin{minipage}[t][6.3cm][t]{1.0\textwidth} % Adjust height [6.3cm] as required for separation of bio photos.
  \begin{wrapfigure}{L}{0.25\textwidth}
    \includegraphics[width=0.25\textwidth]{john_smith.eps}
  \end{wrapfigure}
  \noindent
  {\bfseries John Smith} received his BSc (Mathematics) in 2000 from The University of Maryland. His research interests include lasers and optics.
\end{minipage}
\begin{minipage}{1.0\textwidth}
  \begin{wrapfigure}{L}{0.25\textwidth}
    \includegraphics[width=0.25\textwidth]{alice_smith.eps}
  \end{wrapfigure}
  \noindent
  {\bfseries Alice Smith} also received her BSc (Mathematics) in 2000 from The University of Maryland. Her research interests also include lasers and optics.
\end{minipage}
\endgroup
}{}

\end{document}